# Clustering SPIRES with EqRank

G. B. Pivovarov*
*Institute for Nuclear Research,*
*Moscow, 117312 Russia*

S. E. Trunov†
*SEUS,*
*Bol'shoi Trekhsvyatitel'skii per. 2,*
*Moscow, 109028 Russia*
(Dated: January 8, 2005)

SPIRES is the largest database of scientific papers in the subject field of high energy and nuclear physics. It contains information on the citation graph of more than half a million of papers (vertexes of the citation graph). We outline the EqRank algorithm designed to cluster vertexes of directed graphs, and present the results of EqRank application to the SPIRES citation graph. The hierarchical clustering of SPIRES yielded by EqRank is used to set up a web service, which is also outlined.

PACS numbers:

## I. INTRODUCTION

Internet and www facilitate emergence of Knowledge Networks [1]. The papers are the information units of the network, and the references to other papers are the links of the network. In the organization of the scientific knowledge, there is a hierarchy of transforming and developing scientific themes. While the presence of themes in the organization of knowledge can hardly be questioned, it is a nontrivial problem to reveal this hidden hierarchy with an algorithm analyzing the network structure. Attempts on finding such algorithms had been initiated in the seventies [2, 3], and continued until now [4, 5, 6]. These algorithms attempt to find themes as clusters of papers. As a rule, they involve a number of free parameters (e.g., number of clusters, number of hierarchy levels, citation thresholds, etc.). The results yielded by the algorithms depend strongly on the values of the parameters. Recently, a new algorithm, EqRank has been suggested [7]. Essentially, it does not involve free parameters, and reveals the hierarchy of themes objectively present in the structure of knowledge network. EqRank has been applied to the hep-th citation graph (see the results in [7], and at http://hepstructure.inr.ac.ru/hep-th). In this paper, we outline the EqRank algorithm, and describe the results of its new application. This new application is the first application of EqRank to a massive dataset. The citation graph of the SPIRES database has been chosen as the input data for this application. As a result, a four level classification of the papers from SPIRES has been obtained, and the papers have been indexed in this classification scheme. A web service based on this classification and indexing has been set up. Its alpha version can be accessed at http://hepstructure.inr.ac.ru/index_new2.htm. In the next section, we outline the EqRank algorithm; in the third section, we describe application of EqRank to clustering papers from SPIRES; in the fourth section, we describe the hepstructure web service; in the last section, we discuss the prospects for EqRank applications.

## II. EQRANK

The Eqrank algorithm can be applied to any directed graph. We explain the idea behind the EqRank algorithm using the terms natural to a citation graph. Assume that we have learned somehow the way to compute the local hub paper LH(p) for each paper p. LH(p) cites p and is the most representative paper among the papers developing the ideas of p. The mapping LH(p) generates the trajectory (p, LH(p), LH(LH(p)), ...). The trajectory starts at the paper p, and ends at the paper RH(p), which has no citations. We call the end point of the trajectory the root hub of p. Let us introduce a partition of the set of papers into modern themes. Two papers p and q are in the same modern theme if they share the root hub, RH(p)=RH(q). A modern theme is formed with the papers that share a common resulting paper, the root hub, which is the paper underscoring the present state of the modern theme. In complete analogy, if we know for each paper p the local authority LA(p) that is the paper cited by p, and is the most representative paper among the papers on which p is based, we determine the partition of the set of papers onto classic themes. Each paper in a classic theme has one and the same paper as its root authority. Frequently, a root authority is a seminal paper initiating a new direction of research. We call simply the themes the elements of the partition yielded by intersection of the hub partition and the authority partition. All the papers of a theme have one and the same root hub and authority papers. As soon as we de-fined themes, we can shrink them to vertexes of a reduced graph and repeat the procedure, which ultimately yields a hierarchical clustering of the papers (after the second iteration, the themes, obtained on the first iteration become subthemes of a larger theme, etc.). What remains to be explained is how to find local authorities and hubs. In EqRank, the links (references) are weighted by their co-citation [2], and the most weighted link leads from a paper to its local authority, and, after inversion of link directions, to its local hub. The above heuristic description of EqRank ignores important details, for them, see [7]. But the general idea behind EqRank should be clear: we traverse the most important references from each paper down to the root authority, and the most important citations up to the root hub. The roots characterize a paper. Two papers characterized by the same roots belong to one and the same theme. In [7], a mathematical consideration is presented, which demonstrates that the partition of EqRank is in certain sense the only natural partition onto themes. Let us now turn to the first application of EqRank to a massive dataset, the dataset of SPIRES.

---

*Electronic address: gbpivo@ms2.inr.ac.ru
†Electronic address: trunov@msk.seus.ru

## III. CLUSTERING SPIRES

SPIRES is a database of scientific papers in the subject field of high energy and nuclear physics. Via a web interface at http://www.slac.stanford.edu/spires/hep/, it gives a possibility to search for papers; each paper yielded by a search comes with a link to the list of its references to other papers. Therefore, SPIRES contains information about the citation graph (the graph whose vertexes are the papers, and directed links, the references). We have been harvesting SPIRES and its mirrors around the globe with a robot collecting the information on the citation graph. (Here we wish to thank the people behind the SPIRES mirrors for their collaboration). The obtained citation graph has been processed by EqRank. The SPIRES citation graph used as EqRank input consisted of 1 053 194 vertexes and 6 270 238 links. Among the vertexes, 558 229 correspond to the papers from SPIRES, and 494 965, to the papers outside the SPIRES, but cited from it. EqRank was applied to the largest connected component of this graph, which contains 822 622 vertexes, among which 330 783 papers are inside SPIRES, and 491 835, outside. Application of EqRank has yielded a four-level classification scheme. The papers are on the ground level. They are clustered into 885 themes of the first level. The themes of the first level are clustered into 254 themes of the second level, which are clustered into 52 themes of the third level, and, finally, on the top, fourth level, we have only 6 themes. We stress that EqRank has as its input only the citation graph, and neither the number of levels, nor the number of themes on each level are preset. Instead, the set of themes and indexing of papers with the themes are computed by EqRank from scratch. The natural question to address at this stage is on how to estimate the quality of the obtained classification. It is a nontrivial issue, since there is no any obvious alternative classification to compare with (say, if we try to compare with classification by PACS numbers, they are available for only a fraction of the papers from SPIRES). We can use an imitation of expert estimate, finding in www lists of papers on specific scientific topics recognized by scientific community, and comparing them with the corresponding lists generated by EqRank. In a few cases where we were able to make such an imitation, the comparison is always favorable for EqRank: it seems that EqRank includes in the themes all the papers pointed out by experts, and adds some extra papers overlooked by the experts. We stress that we have only preliminary observations on the quality of the clustering yielded by EqRank, but these observations are very promising.

## IV. WEB SERVICE

Results yielded by EqRank have been used to set up a web service. In the service, papers can be searched for by key words and/or author names, similar to the search possibilities provided by SPIRES, but the classification is supported in forming the responses to the inquiries: e.g., for a paper found, its tree of themes is also provided, so one knows what are the themes the paper belongs to. It is also possible to browse the theme tree. Another supported feature is the availability of lists of hub and authority citations for any theme (including papers, which are the themes of the ground level). In this way, a newcomer locates all the papers relevant for studying a new theme in several clicks. Alfa version of the service is available at http://hepstructure.inr.ac.ru/index_new2.htm.

## V. DISCUSSION

The EqRank classification of SPIRES is the first publicly available automatically generated classification of a large database of scientific papers. (To the best of our knowledge, there are only projects of this sort, not the results. Among the projects are "Natural Communities in Large Linked Networks" at http://citeseer.ist.psu.edu/hopcroft03natural.html, and "Clustering Methods Based on Minimum-Cut Trees" at http://citeseer.ist.psu.edu/607238.html). What are possible uses of such classification? First, it facilitates search. But there is less obvious use: global analysis of a knowledge network may reveal that detached researchers work on a same theme, and intensify collaboration. Also, updating classification for fresh papers may reveal emergent themes, and help to decide on direction of research. Among the possible further applications of EqRank is classification of the papers from CiteSeer database, and from other databases of scientific papers.

This work was supported in part by RFBR grant no. 04-07-90132.